# Particle acceleration by a solar flare termination shock

**Authors:** Bin Chen[1*], Timothy S. Bastian[2], Chengcai Shen[1], Dale E. Gary[3], Säm Krucker[4,5], Lindsay Glesener[4,6]

**Affiliations:**

[1]Harvard-Smithsonian Center for Astrophysics, 60 Garden St, Cambridge, MA 02138, USA

[2]National Radio Astronomy Observatory, 520 Edgemont Rd, Charlottesville, VA 22903, USA

[3]New Jersey Institute of Technology, 323 Martin Luther King Blvd, Newark, NJ 07102, USA

[4]University of California, Berkeley, 7 Gauss Way, Berkeley, CA 94720, USA

[5]University of Applied Sciences and Arts Northwestern Switzerland, Bahnhofstrasse 6, 5210 Windisch, Switzerland

[6]University of Minnesota, Twin Cities, 116 Church Street S.E., Minneapolis, MN 55455

*Correspondence to: bin.chen@njit.edu

**Abstract**: Solar flares—the most powerful explosions in the solar system—are also efficient particle accelerators, capable of energizing a large number of charged particles to relativistic speeds. A termination shock is often invoked in the standard model of solar flares as a possible driver for particle acceleration, yet its existence and role have remained controversial. We present observations of a solar flare termination shock and trace its morphology and dynamics using high-cadence radio imaging spectroscopy. We show that a disruption of the shock coincides with an abrupt reduction of the energetic electron population. The observed properties of the shock are well-reproduced by simulations. These results strongly suggest that a termination shock is responsible, at least in part, for accelerating energetic electrons in solar flares.

**One Sentence Summary:** A termination shock is captured in action during a solar flare using radio observations, which show that it is a source of energetic electrons.

**Main Text:**

The acceleration of charged particles to high energies occurs throughout the Universe. Understanding the physical mechanisms is a fundamental topic in many space, astrophysical, and laboratory contexts that involve magnetized plasma (*1*). For solar flares and the often associated coronal mass ejections (CMEs), it is generally accepted that fast magnetic reconnection—the sudden reconfiguration of the magnetic field topology and the associated magnetic energy release—serves as the central engine driving these powerful explosions. However, the mechanism for converting the released magnetic energy into the kinetic energy in accelerated particles has remained uncertain (*2*, *3*). Competing mechanisms include acceleration by the reconnection current sheet, turbulence, and shocks (*2–5*).

Of possible interest in this regard is the termination shock (TS), produced by super-magnetosonic reconnection outflows impinging upon dense, closed magnetic loops in a cusp-shaped reconnection geometry (*6*). Although often invoked in the standard picture of solar flares (*7*, *8*) and predicted in numerical simulations (*6*, *9–11*), its presence has yet to be firmly established observationally and, because of the paucity of direct observational evidence, its role as a possible particle accelerator has received limited attention (*2*, *3*). Previous reports of coronal hard X-ray (HXR) sources in some flares have shown convincing evidence of the presence of accelerated electrons at or above the top of flare loops (referred to as the "loop-top" hereafter, or LT) (*7*, *12*), where a TS is presumably located. The often cited observational evidence for a solar flare TS has been certain radio sources showing spectroscopic features similar to solar type II radio bursts (radio emission associated with propagating shocks in the outer corona), but with small drifts in their emission frequency as a function of time, which implies a standing shock wave (*13–17*). However, because of the limited spectral imaging capabilities of the previous observations, none of these have shown direct signatures of the TS in terms of its characteristic morphology and dynamics, as well as a clear relation to the reconnection outflows, so that a definitive association with a TS could be demonstrated.

We present observations of a TS in an eruptive solar flare using the Karl G. Jansky Very Large Array (VLA). This eruption occurred close to the east limb of the Sun (Fig. 1A), producing a fast white light CME [~1000 km s$^{-1}$; observed by the Large Angle and Spectrometric Coronagraph Experiment (LASCO)] and a C1.9-class long-duration flare (*18*). It displayed a cusp-shaped reconnection geometry typical of the standard scenario of eruptive solar flares (*7*, *8*) in which the eruption outward into the upper corona stretches magnetic field lines behind it and induces a vertical current sheet where magnetic reconnection occurs. The reconnected field lines below the reconnection site are pulled downward by magnetic tension to form an arcade of magnetic loops anchored at the solar surface. The arcade of reconnected loops subsequently fills with hot plasma and becomes bright in extreme ultraviolet (EUV) and soft X-ray (SXR) wavelengths. For this event, the eruption, the current-sheet-like structure, and the cusp-shaped magnetic loops are all clearly visible in EUV and SXR passbands that are sensitive to plasma hotter than ~2 MK (Fig. 1A). A non-thermal HXR source appears at the LT during the rise phase of the flare, indicating the presence of accelerated electrons at this location (Fig. 1B).

VLA images at 1 to 1.8 GHz show a localized radio source nearly cospatial with the HXR LT source, in addition to two other sources located near the conjugate magnetic footpoints (FPs) of the flaring loops (Figs. 1B and S1). The VLA's simultaneous high spectral and temporal resolution (1 MHz and 50 ms, respectively, enabling high-cadence radio imaging spectroscopy) reveals the highly dynamic and fragmented nature of this LT radio source. It consists of thousands of short-lived (<50 ms) and narrow-frequency-bandwidth (with spectral width $\delta v / v \approx 2\%$) brightenings (Fig. 2D-E; *19*). These observations strongly imply that many short-lived emission events, which we term stochastic radio spikes, are occurring at the LT, which, as we will demonstrate, are associated with a dynamic TS.

Difference imaging in the EUV 94 Å passband of the Atmospheric Imaging Assembly (AIA) aboard the Solar Dynamics Observatory (*20*) reveals that many recurring plasma downflows (PDs) stream rapidly (at ~550 km/s in projection) along the current sheet from the reconnection site downward to the flaring, reconnected loops. They end near the same location as the LT radio and HXR sources (Fig. 2C). These fast PDs are thought to be associated with magnetic structures embedded in reconnection outflows, probably in the form of rapidly contracting magnetic loops (*12*). The relative locations of the PDs and the radio/HXR LT sources agree very well with the scenario in which a TS forms at the ending fronts of fast reconnection outflows and drives particle acceleration.

The most direct observational evidence of the TS comes from the instantaneous spatial distribution of the myriad radio spikes at different frequencies, which forms a narrow surface at the LT region (Fig. 3A). The location and morphology of this surface closely resemble those of a TS as predicted in numerical simulations when viewed edge on (*6*, *9–11*; see also Fig. 3B). The coronal HXR source is located slightly below this surface and evolves coherently with it (Figs. 3A and S4), suggesting that this is non-thermal emission from accelerated electrons confined in the shock downstream region, possibly due to strong pitch-angle scattering and/or magnetic trapping in the turbulent environment (*21*). The TS is probably a weak quasi-perpendicular fast-mode shock based on multiple lines of evidence (*19*). A Mach number of $M \approx 1.5$ can be inferred based on the interpretation of the split-band feature seen in the radio dynamic spectrum [marked HF (high-frequency) and LF (low-frequency) in Fig. 2D; *19*].

The TS front, as outlined by the radio spikes, reacts dynamically to the arrival of the fast PDs. Some PDs cause only partial disruption of the TS front, and the shock is quickly restored to its original state. Some other PDs, however, cause a major disruption of the TS. This process starts with the quasi-flat TS front being first driven concave-downward by a PD, followed by a breakup of the TS (Figs. 3D and S4). To understand the dynamic nature of the TS, we use a magnetohydrodynamics (MHD) model to simulate magnetic reconnection in a standard flare geometry based on physical values constrained by the observations (*19*). The model shows that reconnection outflows with super-magnetosonic speeds can produce a TS in the LT region, and the observed morphology and dynamics of the TS are well-reproduced by the simulations (Figs. 3B and E; Movie S1). In the simulations, the observed PDs correspond to magnetic structures formed because of instabilities in the reconnection current sheet, which may facilitate the efficiency of the magnetic energy release that powers solar flares (*22*).

During the largest disruption of the TS, the intensities of all of the three widely-separated radio sources decrease simultaneously. The HXR flux above 15 keV is also abruptly reduced, whereas the SXR flux (<12 keV) is largely unaffected, which is consistent with a temporarily softened X-ray photon spectrum (Fig. 4). Both phenomena suggest a temporary decrease of the number of energetic electrons. By fitting the observed X-ray spectrum using an isothermal plasma plus a nonthermal electron distribution with a power-law form, we confirmed that energetic electrons were much less abundant during the shock destruction: the total number of >18 keV electrons was reduced by ~62% (*19*). This is strong evidence that the TS plays a key role in accelerating the energetic electrons.

An important question is what emission mechanism is responsible for the multitudes of narrowband stochastic radio spikes at the TS. An attractive possibility is linear mode conversion of Langmuir waves on small-scale density fluctuations (*23, 24*), a mechanism that has been explored in the context of radio bursts in the solar corona, in Earth's foreshock region, and near the heliospheric TS (*24–26*). This mechanism requires both a source of Langmuir waves and the presence of small-scale density fluctuations. We suggest that electrons are accelerated in the turbulent plasma environment at the TS (*5, 16, 27, 28*), an assumption supported by the HXR source at the LT (Fig. 3C). These accelerated electrons are unstable to the production of Langmuir waves, which impinge on the small-scale density fluctuations associated with the turbulent medium and convert to electromagnetic waves near the local plasma frequency $\nu_{pe} = (e^2 n_e / \pi m_e)^{1/2} \approx 8980\sqrt{n_e}$ Hz, where $n_e$ is the electron density (*23, 24*). The frequency range of 1 to 1.8 GHz over which the spike bursts appear then implies a density range of $n_e \approx 1.2 - 4 \times 10^{10}$ cm$^{-3}$, which is consistent with that from the X-ray spectral analysis (*19*). The level of the density fluctuations $\delta n_e / n_e$ is related to the observed spike bandwidths as $\delta n_e / n_e \approx 2 \delta \nu / \nu$, which is relatively small (4%). The spatial scales of the density fluctuations are also small, a few hundred kilometers at the maximum (*19*).

A major theoretical concern regarding electron acceleration by a fast-mode quasi-perpendicular shock (as for the case of a TS) has been the injection problem: Electrons need to cross the shock front multiple times and/or be pre-accelerated to suprathermal energies in order to gain energy efficiently (*2, 3, 27*). Our observations show strong evidence for the existence of many small-scale low-amplitude fluctuations at the TS front, which may serve as scattering agents that cause repeated passage of the electrons across the shock (*5, 27–30*). In addition, the nonthermal electron population is reduced but not eliminated during the TS disruption (Figs. 4 and S5), which implies that electrons may have been pre-accelerated before they reach the shock, possibly at or near the reconnection site (*2–4, 12*). Both signatures may contribute to resolving the injection problem.

By confirming the existence of the previously controversial solar flare TS and providing strong evidence for it being a particle accelerator, we have obtained new insights into the long-standing problem of particle acceleration in solar flares.

**Acknowledgments:** The authors would like to thank the National Radio Astronomy Observatory (NRAO) staff for their support, and thank Ed DeLuca, Kathy Reeves, Hui Tian, Jun Lin, Alexander Warmuth, Hugh Hudson, Amir Caspi, Fan Guo, Gelu Nita, Gregory Fleishman, and Xuening Bai for their helpful discussions. Stephen Bourke and Gregg Hallinan are acknowledged for making their fast radio imaging software AIPSLITE available for BC. Anne K. Tolbert and Richard Schwartz are thanked for their help in providing the Fermi/GBM detector response for this flare. Zhitao Wang and Marc DeRosa are thanked for their help on the PFSS extrapolations. The NRAO is a facility of the National Science Foundation (NSF) operated under cooperative agreement by Associated Universities, Inc. The VLA data can be accessed at https://archive.nrao.edu/archive/advquery.jsp, using the observing time as the search criteria. The Solar Dynamics Observatory/AIA, RHESSI, Hinode/XRT, and SOHO/LASCO data are all available through the Virtual Solar Observatory (http://sdac.virtualsolar.org/cgi/search). BC acknowledges support by NASA under contract SP02H1701R from Lockheed-Martin to SAO and contract NNM07AB07C to the Smithsonian Astrophysical Observatory, and by the NASA Living-with-a-Star Jack Eddy Fellowship (administered by the University Corporation for Atmospheric Research). CS's work was supported by NSF SHINE grants AGS-1156076 and AGS-1358342 to SAO. DG acknowledges support from NSF grant AST-1312802 and NASA grant NNX14AK66G to New Jersey Institute of Technology. SK and LG are supported by NASA contract NAS598033 for the RHESSI spacecraft.

**Supplementary Materials:**

Materials and Methods

Figures S1-S6

Movie S1

References (*31-56*)

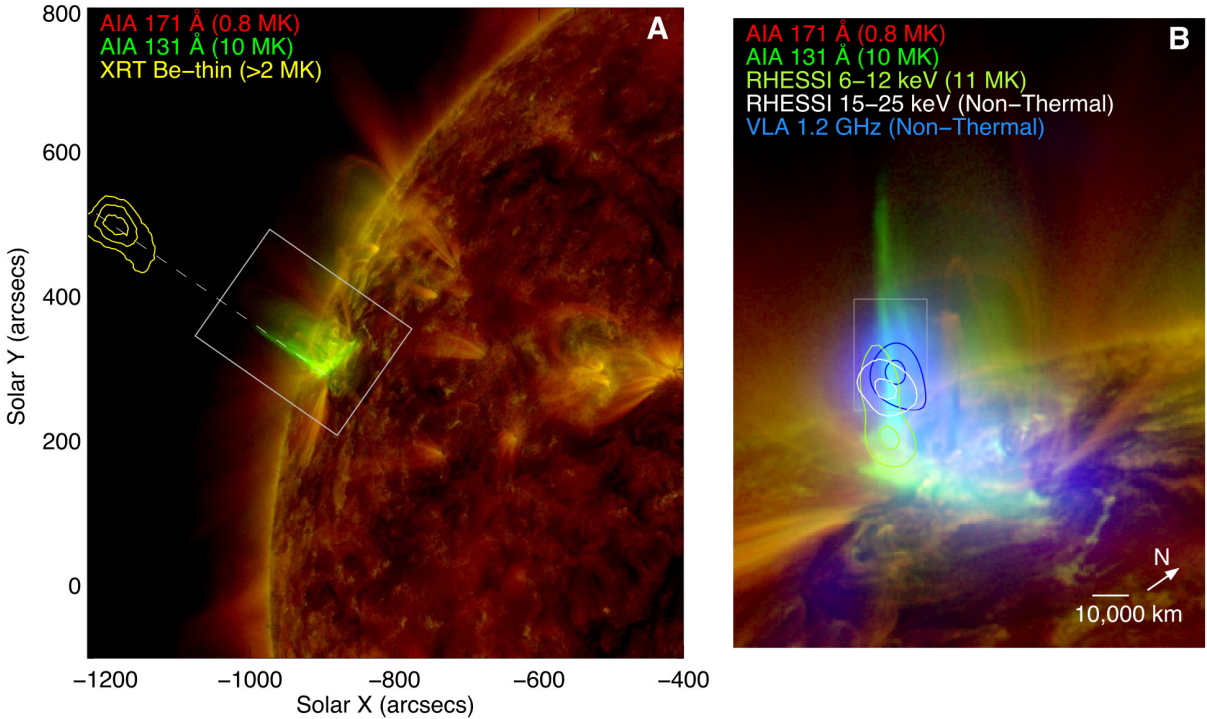

**Fig. 1. Solar flare seen in multiple wavelengths**. (**A**) The eruptive flare observed in EUV and X-ray wavelengths by the Atmospheric Imaging Assembly 171 Å (red), X-Ray Telescope (XRT; aboard Hinode) Be-thin (yellow contours, showing the eruption), and AIA 131 Å (green, showing the newly-reconnected flare loops) passbands, which are respectively sensitive to plasma temperatures of 0.8 MK, >2 MK, and 10 MK. (**B**) Closer view of the flaring region (box in **A**, rotated clockwise to an upright orientation). A radio source (blue; at 1.2 GHz) is observed at the top of hot flaring loops (~10 MK), which is nearly cospatial with a non-thermal HXR source (white contours; at 15–25 keV) seen by the Reuven Ramaty High Energy Solar Spectroscopic Imager (RHESSI).

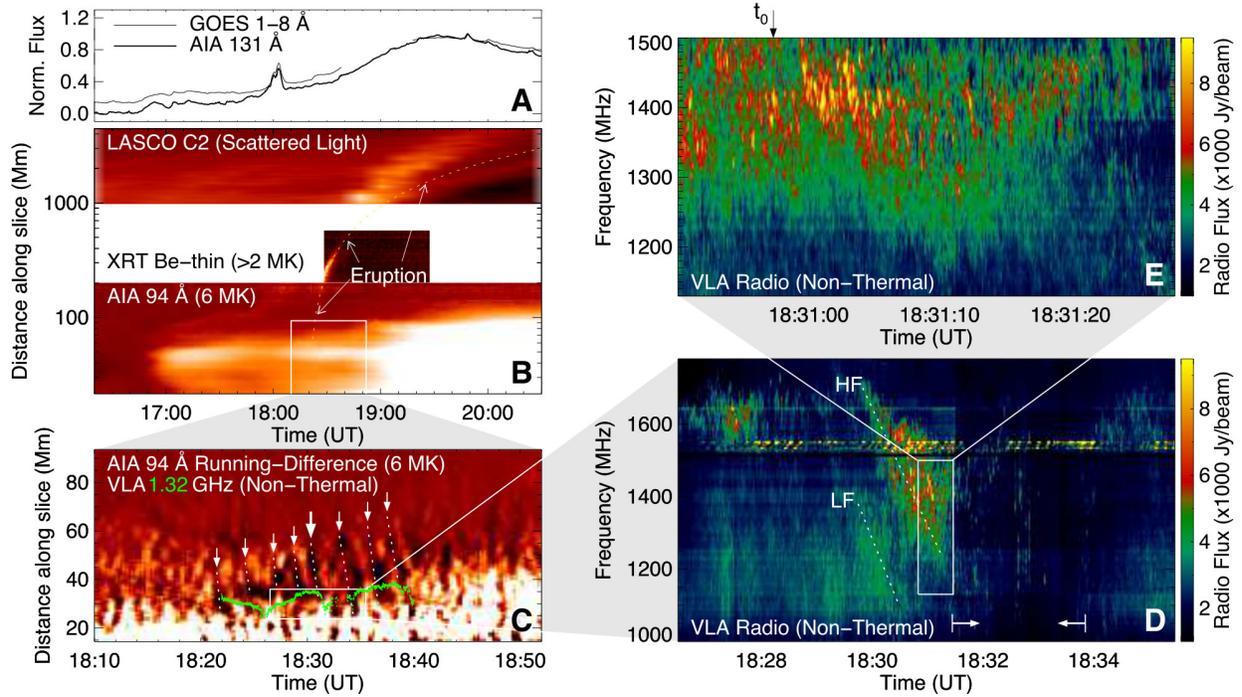

**Fig. 2. Radio emission at the front of fast reconnection outflows.** (**A**) EUV and X-ray light curves showing the time history of the radiating hot plasmas (≥ 10 MK) of the flare event. The TS is observed during the flare rise phase. (**B**) Time-distance plot of the EUV, X-ray, and white light intensities showing the evolution of the eruption and the underlying flare loops, obtained at the slice in Fig. 1A (dashed line). (**C**) Running-difference space-time plot of EUV 94 Å zoomed in to the LT region. A series of fast PDs are visible as features with a negative slope (indicated by white arrows; the PD associated with the TS disruption in Fig. 3 is marked by a thick arrow). The stochastic spike bursts are located near the endpoint of these PDs (green dots). Its spectro-temporal intensity variation is shown in the spatially-resolved, or "vector" dynamic spectrum of (**D**) and (**E**), manifesting as many short-lived, narrow-frequency-bandwidth radio bursts. Two dotted lines in (**D**) mark the split-band feature (HF and LF denote the high- and low-frequency branch, respectively). A pair of arrows brackets a period when the TS experiences a major disruption starting from 18:31:27 UT.

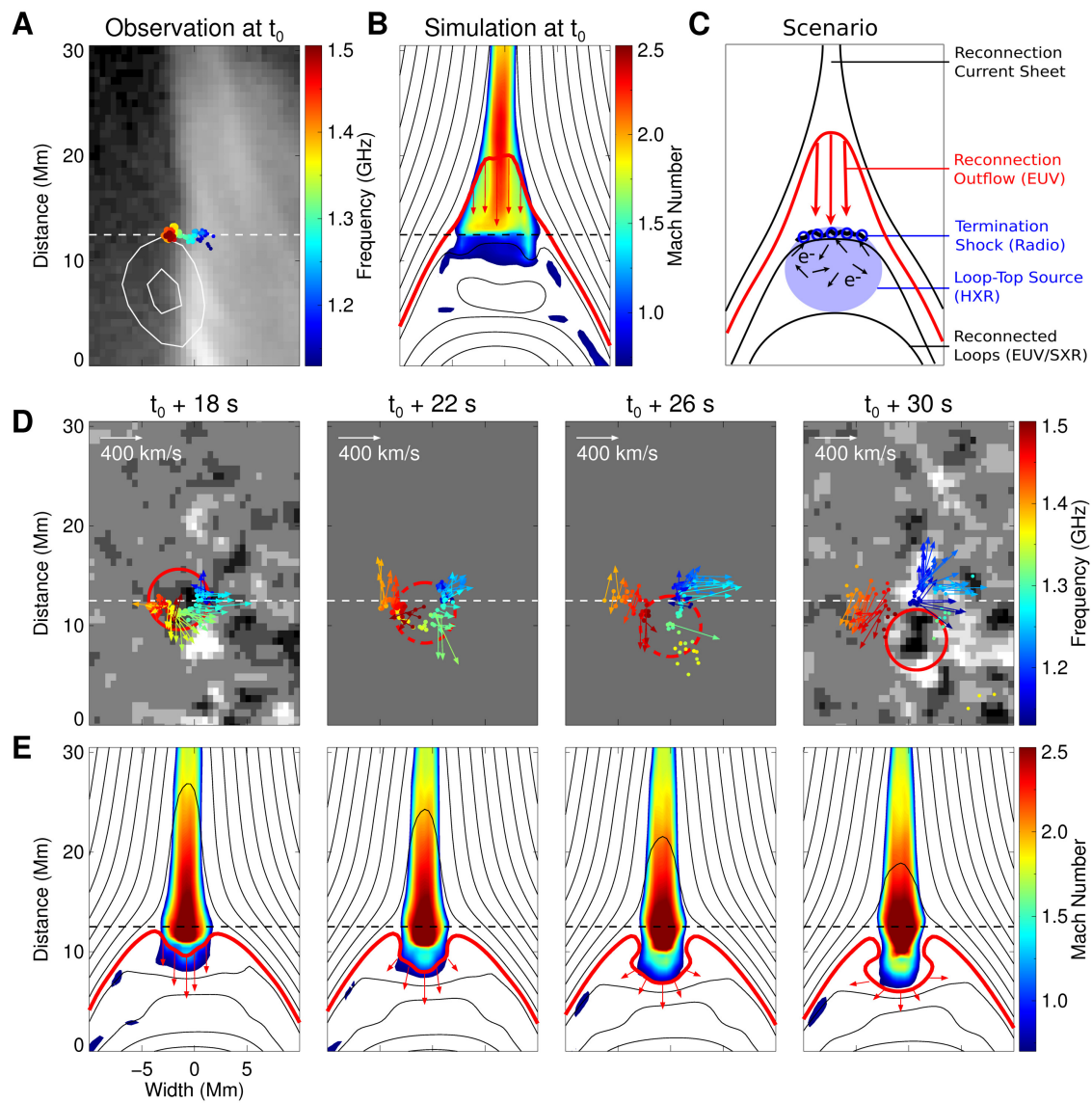

**Fig. 3. Observation and simulation of the dynamic termination shock.** (**A**) A closer view of the LT region (white box in Fig. 1B) at 18:30:57 UT (denoted as $t_0$ in Fig. 2E). The TS appears as a dynamic surface delineated by the many unresolved radio sources, each of which corresponds to a radio spike in the dynamic spectrum at a given time and frequency (colored dots indicate their centroid location). White contours show the coronal HXR source at 15–25 keV. The grayscale background is the AIA 94 Å intensity. (**B**) The TS is seen in the MHD simulation as a sharp layer of velocity discontinuity at the LT. The fast-mode magnetosonic Mach number is shown in color, overlaid with magnetic field lines. (**C**) Physical scenario of emission processes near the TS. Radio spikes are emitted as accelerated electrons impinge density fluctuations at the shock (blue circles). These electrons also produce a HXR source in the shock downstream region (blue shadowed region). (**D**, **E**) Observation and simulation of the TS disruption. A fast PD identified in the AIA 94 Å running-difference images (red circles) arrives at the TS at ~18:31:15 UT ($t_0 + 18$ s) and disrupts the shock, which appears in the simulation as a rapidly contracting magnetic loop (red curve). Arrows show the velocity vectors.

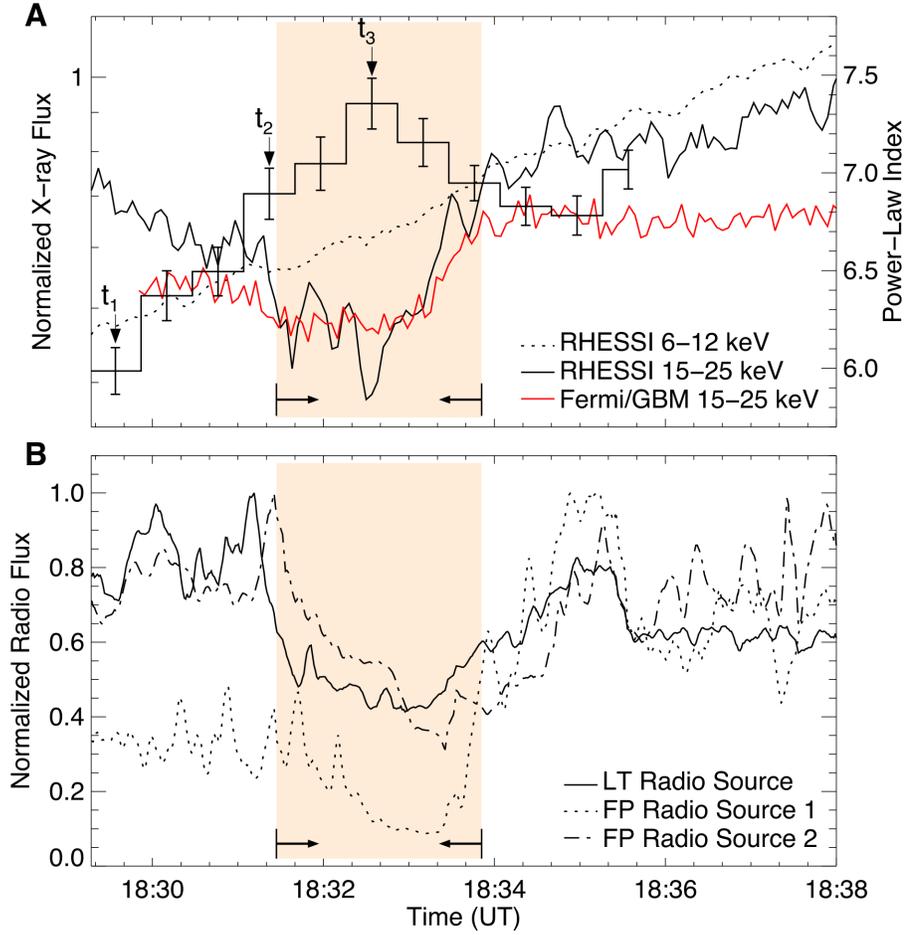

**Fig. 4. Reduction of X-ray and radio flux during shock disruption.** (**A**) Temporal evolution of total X-ray photon counts (curves) and power-law index of the X-ray photon spectrum (histogram). Examples of the observed and fitted X-ray spectra are shown in Fig. S5 for selected times before and during the shock disruption (marked as $t_1, t_2$, and $t_3$). (**B**) Evolution of the spatially-resolved radio flux of the LT source and the two FP sources, showing a co-temporal radio flux reduction (see also fig. S1C). Arrows bracket a period when the TS experiences a major disruption starting from 18:31:27 UT (or $t_0 + 30$ s, corresponding to the last panel of Fig. 3D).

# Supplementary Materials for

## Particle acceleration by a solar flare termination shock

Bin Chen*, Timothy S. Bastian, Chengcai Shen, Dale E. Gary, Säm Krucker, Lindsay Glesener

*Correspondence author. E-mail: bin.chen@njit.edu

**This PDF file includes:**

    Materials and Methods
    Figs. S1 to S6
    Caption for Movie S1

**Other Supplementary Materials for this manuscript includes the following:**

    Movie S1

**Materials and Methods**

S1. Radio Data and Analysis

The VLA is a general-purpose radio telescope array consisting of 27 antennas, each with a diameter of 25 m. It operates over a wide frequency range of 58 MHz to 50 GHz. In 2011, the National Radio Astronomy Observatory completed a major upgrade of the VLA with state-of-art receivers and electronics, which greatly enhanced its capabilities (*31*). Provisions were made with the upgraded VLA to enable solar observing mode, which was successfully commissioned and tested in late 2011 (*32*). The relatively large instantaneous bandwidths for imaging, the large number of spectral channels, and the fast sampling times available from the upgraded VLA enable the new observing technique of radio dynamic imaging spectroscopy.

We performed the VLA observation of the Sun on 2012 March 3 from 17:39:15 UT to 21:44:54 UT. The VLA was in its C configuration, for which the longest baseline is 3 km. The observation was made in two circular polarizations, with 1024 spectral channels of 1 MHz bandwidth, covering the 1–2 GHz frequency band (15–30 cm in wavelength). The temporal cadence was 50 ms. Fifteen VLA antennas were used for observing due to the limited data throughput available at that time. The angular resolution of the radio images was determined by the full-width at half maximum (FWHM) of the synthesized beam, which was 20.5"×17.3" at 1.5 GHz (and scales proportionally to wavelength) at the time of the observation.

The VLA observation covered the eruptive solar flare event from its initiation to well past the flare maximum. During the SXR rise phase of this event, multitudes of radio bursts were recorded consisting of complex spectro-temporal substructures. The observed radio emission is dominated by three spatially-distinct sources: one located at the LT and two others at the conjugate FPs of the flaring loops (Figs. 1B and S1). Radio dynamic imaging spectroscopy allowed imaging at all available spectral channels with ultra-high temporal cadence in both right- and left-hand-circular polarizations (RCP and LCP), resulting in a high-resolution 4.5-dimensional image cube (two dimensions in space, one in frequency, and one in time for each polarization). We used a fast synthesis imaging software package, AIPSLITE (*33*), to produce such an image cube covering the entire event. By selecting a spatial region in the image cube containing only the source of interest, the intensity variation within the selected region can be obtained as a function of time and frequency. In this way, a "vector" dynamic spectrum is derived for each radio source, which isolates the spectro-temporal information intrinsic to the source itself (Fig. S1) from the contributions of sources at other locations. The LT source is RCP, and appears in the vector dynamic spectrum as an ensemble of many short-lived and narrow-band stochastic spike bursts (labeled as "LT Spikes"). The two FP sources appear as quasi-periodic pulsations with different frequency, bandwidth, and pulsation periods (labeled as "FP Source 1" and "FP Source 2"). They are polarized in opposite senses (RCP and LCP for FP

Sources 1 and 2, respectively), which are radio emission (in the sense of the ordinary, or o-mode electromagnetic radiation) associated with conjugate FPs of the flare loops with opposite magnetic polarities. All three sources are excited by non-thermal electrons that are accelerated in the energy release region, but their emission processes and/or source environments are different. The same event was also captured by the Expanded Owens Valley Solar Array (EOVSA) Subsystem Testbed (EST) with a higher time resolution (20 ms) but with very limited spatial information. The stochastic spike bursts are also present in EST's cross-power dynamic spectrum (resulting from the cross-correlation of radio intensities measured at two spatially-separated antennas), with identical spectral and temporal characteristics (Fig. S2). However, these bursts are difficult to identify in the EST's total-power dynamic spectrum which have no spatial information about the source and therefore includes the contributions from the co-temporal presence of strong radio sources at the FPs. Because all previous studies that attempt to identify radio spectroscopic features of the TS have relied solely on total-power dynamic spectra, this could be one important reason for the scarceness of such reports.

The broadband, multi-frequency imaging capability provided by the VLA is crucial in obtaining the spatial distribution of the radio spike centroids at different frequencies, which effectively tracks a large number of small-scale structures on the shock with different densities, and hence, maps the shock surface (Fig. 3). The spatial location of each radio spike in the dynamic spectrum was obtained by finding the centroid of the radio image at the time and frequency of the corresponding spike. The positional accuracy of the derived centroids is $\sim 1-2"$ for the majority of the spikes, estimated using the synthesis beam size ($\sim 15-30"$) scaled by the signal-to-noise ratio of the source (>10), which is also consistent with the measured dispersion of the spike centroid locations at a single frequency in a short time interval.

VLA's simultaneous high cadence allows us to track the dynamics of the TS front in detail. Although each individual spike's lifetime is less than one sample time (50 ms), the ensemble location of bursts at each frequency is normally steady within a few seconds, and we take that ensemble to mark a given position in the TS. At any given instant, for every radio source at the shock front, a velocity vector is derived based on its position at the same frequency but 4-s later. This time interval is chosen in order to measure spatial movements well-above the position uncertainty and meanwhile, preserve the coherency of the radio-emitting source. In particular, given an 1" position error for the bright spikes and a 4-s time step, the velocity uncertainty is $\sigma_v \approx 180$ km s$^{-1}$. The velocity vectors shown in Fig. 3 all have values above $\sigma_v$. Another condition necessary for this approach is that the emitting-frequency of the radio source, or the mean density of the source, does not change significantly during the 4-s interval (i.e., relative to the bandwidths of the spikes, which are ~20–50 MHz).

The highly dynamic nature of the TS combined with the density inhomogeneity at the shock surface, give rise to the stochastic appearance of the spike bursts in the vector dynamic spectrum.

It is interesting to note the presence of a type-II-burst-like structure at ~18:31 UT that consists of myriad spike bursts (Figs. 2D and E). This drifting type-II-burst-like structure appears to show both a high- and low-frequency branch (denoted "HF" and "LF" in Fig. 2D) with a frequency ratio $\nu_{HF}/\nu_{LF} \approx 1.3$, which could be interpreted as a signature of band splitting that is sometimes present in type II radio bursts. Such a split-band feature may result from plasma emission on both the upstream and downstream sides of the shock, where a density discontinuity is expected (*13*, *14*, *34*). If this is the case, the density jump across the shock can be derived based on the observed frequency ratio $n_2/n_1 = (\nu_{HF}/\nu_{LF})^2 \approx 1.7$. The shock Mach number can thus be estimated using the Rankine-Hugoniot jump conditions, which give $M \approx 1.5$ in this scenario. Examination of the apparent shock height in the data shows that it is remarkably stable at this time. The frequency drift of the type-II-burst-like structure is therefore likely the result of the dynamic TS moving into a region with a lower ambient density along the arcade top rather than a radial drift.

Independent evidence for a weak shock can be derived using observed parameters in the vicinity of the shock. To form a TS shock in the LT region, the upstream flow speed $v$ should be greater than the local fast-mode magetosonic speed $v_{ms}$. The former can be constrained by the speeds of the fast PDs, measured to be ~550 km s$^{-1}$ in the plane of the sky. Note that this value should be only considered as a lower limit of $v$ due to the projection effect. The latter, $v_{ms}$, is determined by the following plasma parameters: density ($n$), temperature ($T$), and magnetic field strength ($B$). Density $n$ can be obtained directly from the radio emission frequency $\nu$, which gives $1.2-4\times10^{10}$ cm$^{-3}$ for $\nu = 1.0-1.8$ GHz. Temperature $T$ is estimated to be 6–10 MK based on the temperature responses of the AIA 94 and 131 Å passbands at which the flare loops are observed. The magnetic field strength $B$ in the LT region is not directly measured. In order to have $M = v/v_{ms} > 1$, $B$ should be no more than a few tens of Gauss. This is consistent with the values derived using the Potential Field Source Surface (PFSS) model (*35*), which calculates the magnetic fields in the corona based on extrapolations from the magnetic field measurements at the photospheric level.

The spectral properties of the observed radio spikes provide important information on the physical properties of the density fluctuations. Following (*36*), we fit the spectral profile at any 50-ms integration of the vector dynamic spectrum using a multi-Gaussian function plus a flat background. The fitting procedure starts from a set of spikes identified as local maxima in a spectral profile, used as the initial guess for their central frequencies and peak fluxes, and then uses a non-linear least-squares minimization technique [based on the IDL package MPFIT (*37*)] to fit the observed profile. Typical examples of the observed and fitted spectral profile are shown in Fig. S3A, which demonstrate a general agreement between the observed spectra and model fits. The central frequency, peak flux density, and full-width-at-half-maximum (FWHM) of each fitted Gaussian function are used to describe the mean frequency, brightness, and bandwidth of the corresponding radio spike. For a selected region in the vector dynamic spectrum that contains about ten thousand individual spike bursts (from 18:30:20 UT to 18:31:20 UT in 1.13–1.50 GHz), we obtain the following results: (1) The spike bandwidth shows large variations at all times and frequencies, and is mostly well-above 1%, which suggests that it is not a natural bandwidth due to the emission process, but is intrinsic to the stochastic nature of the radio-emitting source inhomogeneities (*38*). (2) The spike bandwidth distribution is asymmetric, which peaks at ~23 MHz (or 2% relative to the observing frequency), with a long tail extending to larger values. (3) The spike bandwidth tends to have less dispersion at larger brightness. These findings are generally consistent with earlier studies of solar spike bursts based on total-power dynamic spectral data (i.e., without spatial information) (*38–41*).

We suggest that the stochastic spike bursts may be emitted by linear mode conversion of Langmuir waves to electromagnetic o-mode radiation on density fluctuations near the TS (*23, 24*). The emission is likely due to fundamental plasma radiation as the spikes are highly polarized. The radiation conversion efficiency is proportional to $L^{-1}$, where $L^{-1}$ is the spatial scale of the relevant fluctuations. It can reach $>10^{-5}$ for small-scale fluctuations of $L<$ few km (*23*). The spectrum of density inhomogeneities in the corona is largely unknown on these small spatial scales. Based on recent imaging spectroscopic observations of type III radio bursts in the corona in this frequency band, however, it is known that the corona is extremely finely structured on spatial scales of order 100 km and less (*32*). Moreover, hybrid simulation results suggested that the plasma environment can be highly turbulent in the vicinity of a quasi-perpendicular shock (as for the case of a TS at the LT region), with the presence of inhomogeneities down to ion inertial scales (~10 m for $n_e = 2\times10^{10}$ cm$^{-3}$) (*30, 42*). In our observations, the upper limit of the spatial scale of the density fluctuations can be constrained by the duration of the stochastic spike bursts via $L \sim v_E \delta\tau_E$, where $v_E$ and $\delta\tau_E$ are the characteristic speed and duration of the exciter. The observed instantaneous duration of the spike bursts $\delta\tau$ has a contribution from both the intrinsic duration of the exciter $\delta\tau_E$ and the emission decay time $\delta\tau_D$: $\delta\tau \approx \delta\tau_E + \delta\tau_D$. We find that the spike bursts are not temporally resolved by either the VLA or EST, with time resolutions of 50 ms and 20 ms, respectively. Taking $v_E$ as the velocity of the fast electron beams that excite Langmuir waves, we have $L<600$ km for $v_E = 0.1c$ (where $c$ is the speed of light) and $\delta\tau_E < \delta\tau < 20$ ms. Note the upper limit of the spatial scale can be much smaller if the emission decay time dominates the observed spike duration. For example, an empirical relation based on a large sample of solar spike bursts from ~0.3 and 1.0 GHz (*43*) implied $\delta\tau_D \approx 6-10$ ms at 1–1.8 GHz, which is comparable to the maximum instantaneous duration of the spike bursts (20 ms, based on the EST data).

S2. X-Ray Data and Analysis

RHESSI is a NASA Small-Explorer Mission launched in 2002, which provides imaging and spectroscopy of the Sun in X-rays and γ-rays from 3 keV to 17 MeV (*44*). The X-ray images are reconstructed using the CLEAN algorithm (*45*) based on RHESSI measurements from the front segments of detectors 3, 5, 6, 7, 8, and 9, with 96-s integration time. The angular resolution of the RHESSI X-ray images is determined by the FWHM of the clean beam, which is $\sim 9''$. The RHESSI X-ray spectra are obtained based on measurements from the front segments of detectors 1, 3, 6, 8, and 9 with an integration time of 36 s. There was a moderate level of pileup (which results from two or more photons arriving close in time and being counted by the electronics as a single photon with their energies summed) for the time intervals used for the RHESSI X-ray spectral analysis (up to 30% at certain energies), which is corrected before the spectral fitting. This event was also recorded by the Gamma-ray Burst Monitor (GBM; aboard Fermi) (*46*), a general-purpose space telescope that observes in X-rays and gamma-rays. Temporally-resolved X-ray spectroscopic data from the most sunward Fermi/GBM detector are used to generate the X-ray light curves and spectra. Similar to the RHESSI results, the Fermi/GBM 15–25 keV HXR light curve shows an intensity reduction (Fig. 4A, red curve) during the major disruption of the TS. The Fermi/GBM X-ray spectra are very similar to those from RHESSI, but with a much higher background. Nevertheless, the power-law index of its X-ray photon spectra also shows a sudden softening during the TS disruption, confirming the RHESSI finding. We will henceforth only show the RHESSI spectral analysis results.

An isothermal plasma plus a non-thermal electron population with a power-law energy distribution

$$f(\varepsilon) = A\varepsilon^{-\delta}(\varepsilon_{lc} < \varepsilon < \varepsilon_{hc}) \qquad (S1)$$

is used to fit the RHESSI X-ray spectra from 5 keV to 22 keV, under the assumption that the emission is produced through a thin-target bremsstrahlung radiation process (*47*). The isothermal spectrum is calculated based on the predictions using the CHIANTI package in the SOLARSOFT IDL distribution. For the power-law non-thermal electron energy distribution, we fix the lower- and higher-energy cutoffs to be $\varepsilon_{lc} = 10$ keV and $\varepsilon_{hc} = 1$ MeV, and treat the normalization factor ($A$) and the power-law index ($\delta$) as free parameters in the fitting process.

The RHESSI X-ray spectra from 5 to 22 keV and the fitting parameters for the three sampled times in Fig. 4A are shown in the Fig. S5. The fitted temperature of the thermal plasma ($T$) is about 11 MK at 18:28 UT, which continuously increases to ~12 MK at 18:36 UT. The emission measure of the thermal plasma is found to be $\xi_T = n^2 V_T \approx 3 \times 10^{47}$ cm$^{-3}$ (where $n$ and $V_T$ are number density and source volume of the thermal plasma, respectively), which also increases slightly with time. Using the FWHM size of the thermal X-ray image at 6–12 keV ($51'' \times 8''$), and assuming an 8" thickness of the thermal source (along the line of sight), the number density of the thermal plasma in the flaring loops is estimated to be $n = (\xi_T / V_T)^{1/2} \approx 1.7 \times 10^{10}$ cm$^{-3}$. Within uncertainties (a factor of a few), this is consistent with the number density derived from the emission frequencies of the stochastic radio spikes based on fundamental plasma radiation ($1.2-4\times10^{10}$ cm$^{-3}$). Note that, for the LT volume, the estimated column density is $N=nd\approx1\times10^{19}$ cm$^{-2}$ (taking the column depth $d\approx8"$), which is insufficient to stop electrons with energy of >10 keV (*47*, *48*). This is consistent with the thin-target assumption for fitting the non-thermal component of the X-ray spectrum, as the LT coronal source is clearly seen to dominate the HXR emission above ~15 keV during the major disruption of the TS from ~18:30 to 18:33 UT (Fig. S4). The insignificance of a HXR chromospheric FP source may be the result of the relatively high collisionality of the coronal source (the electron-electron collisional mean free path for a 15 keV electron in the coronal source is ~50 Mm, comparable to the length of the post-flare loops), which effectively thermalizes most of the non-thermal electrons before they reach the chromosphere (*49*).

The non-thermal electron flux ($NV_{NT}F_e$, in electrons cm$^{-2}$ s$^{-1}$) in the emitting volume $V_{NT}$ above certain energy can be obtained by integrating the fitted non-thermal electron energy spectrum $f(\varepsilon)$ in energy. We find that the total non-thermal electron flux is significantly reduced (e.g., by 62% for >18 keV electrons) during the TS disruption, consistent with the decreasing radio and HXR intensities and the softening X-ray photon spectrum as shown in Fig. 4.

S3. Numerical Simulations

We perform a two-dimensional resistive magnetohydrodynamic (MHD) simulation to investigate the formation and evolution of a TS at the front of fast reconnection outflows during solar eruptions. We use ATHENA, a well-developed grid-based MHD code (*50*), to solve the full resistive MHD equations. As the system can evolve self-consistently, governed by the resistive MHD equations, our numerical experiment is then formulated as an initial and boundary value problem. Our simulation starts with a vertical current sheet in mechanical and thermal equilibrium that separates two regions of the magnetic field with opposite polarity which are line-tied at the lower boundary representing the photosphere (Fig. S6, first column). This configuration has been previously used to model magnetic reconnection in the standard flare geometry and the associated phenomena (i.e., a Kopp-Pneuman configuration (*51*) of two-ribbon flares) following a solar eruption (*51–53*). The boundary conditions are arranged in this way: the right, left, and top sides of the simulation domain are the open boundaries on which the plasma and the magnetic flux are allowed to enter or exit freely. The boundary at the bottom is set to ensure that the magnetic field is line-tied to the photosphere, meaning that these field lines do not move at this boundary.

The simulation is performed on uniformly spaced $400 \times 400$ Cartesian grids. Throughout the simulation region, the resistivity is uniform and corresponds to a magnetic Reynolds number of $1 \times 10^4$. The ohmic heating term is included in the equations governing energy conservation, while the gravity and thermal conduction terms are not included in the current case. Essential physical quantities in the resistive MHD equations, including magnetic field, plasma density, pressure, length, time, and velocity are firstly normalized to dimensionless units for the ease of numerical calculations, and then scaled to physical units according to values constrained by the observations.

The initial equilibrium in the system breaks down as a result of perturbations imposed on the system. Plasma and magnetic fields on the two sides of the current sheet move toward one another gradually, causing the current sheet to become thinner and thinner. Magnetic reconnection commences gradually during this period, which produces two reconnection outflow jets inside the current sheet: one is directed upward, away from the solar surface, and the other downward, toward an arcade of newly-reconnected magnetic loops anchored at the photosphere (Fig. S6, second column). The current sheet continually narrows until it becomes thin enough to develop the tearing mode and other instabilities inside it (*54–56*). As a result, small-scale magnetic structures appear in the current sheet, inducing multiple *X*-shaped magnetic neutral points between each pair of the magnetic islands (Fig. S6, third and fourth columns) where fast magnetic reconnection can occur. The initially quasi-static current sheet then evolves into a non-uniform and dynamic state, when the reconnection outflows become highly intermittent.

The arcade of newly-reconnected, dense magnetic loops acts as an obstacle that stops the high-speed reconnection outflows at the LT. The speed of the reconnection outflows gradually increases as the current sheet continually narrows down. As the outflows become super-magnetosonic, a standing fast-mode shock forms at the LT, exhibiting itself as a TS. The shock can be identified as a sharp transition region where the flow speed abruptly changes from super-magnetosonic to sub-magnetosonic values, as shown in the Mach number plots of Fig. S6B.

The TS firstly appears as a quasi-stationary narrow layer at the LT while the current sheet is still structureless and the reconnection outflows are relatively steady (Fig. S6, second column). Once the instabilities kick in, bursty magnetic reconnections occur in the now highly structured current sheet, resulting in multiple moving magnetic structures along with intermittent reconnection outflows. As they collide on the top of closed magnetic loops, the TS becomes highly dynamic. In particular, the TS can be driven either upward or downward, broadened or narrowed (in its horizontal extension) in response to the arrival of the upstream magnetic structures and fast plasma flows. Its appearance can also vary from a quasi-flat morphology to a curved morphology (Fig. 3). Sometimes the TS signature can temporarily disappear from the LT region, when the instantaneous upstream flow speed becomes sub-magnetosonic and/or the LT region is too turbulent to sustain a well-defined TS during the arrival of one or more magnetic structures. Such a disruption of the TS may lead to the abrupt reduction of the energetic electron population as discussed in the main text, serving as strong evidence for the TS as a particle accelerator.

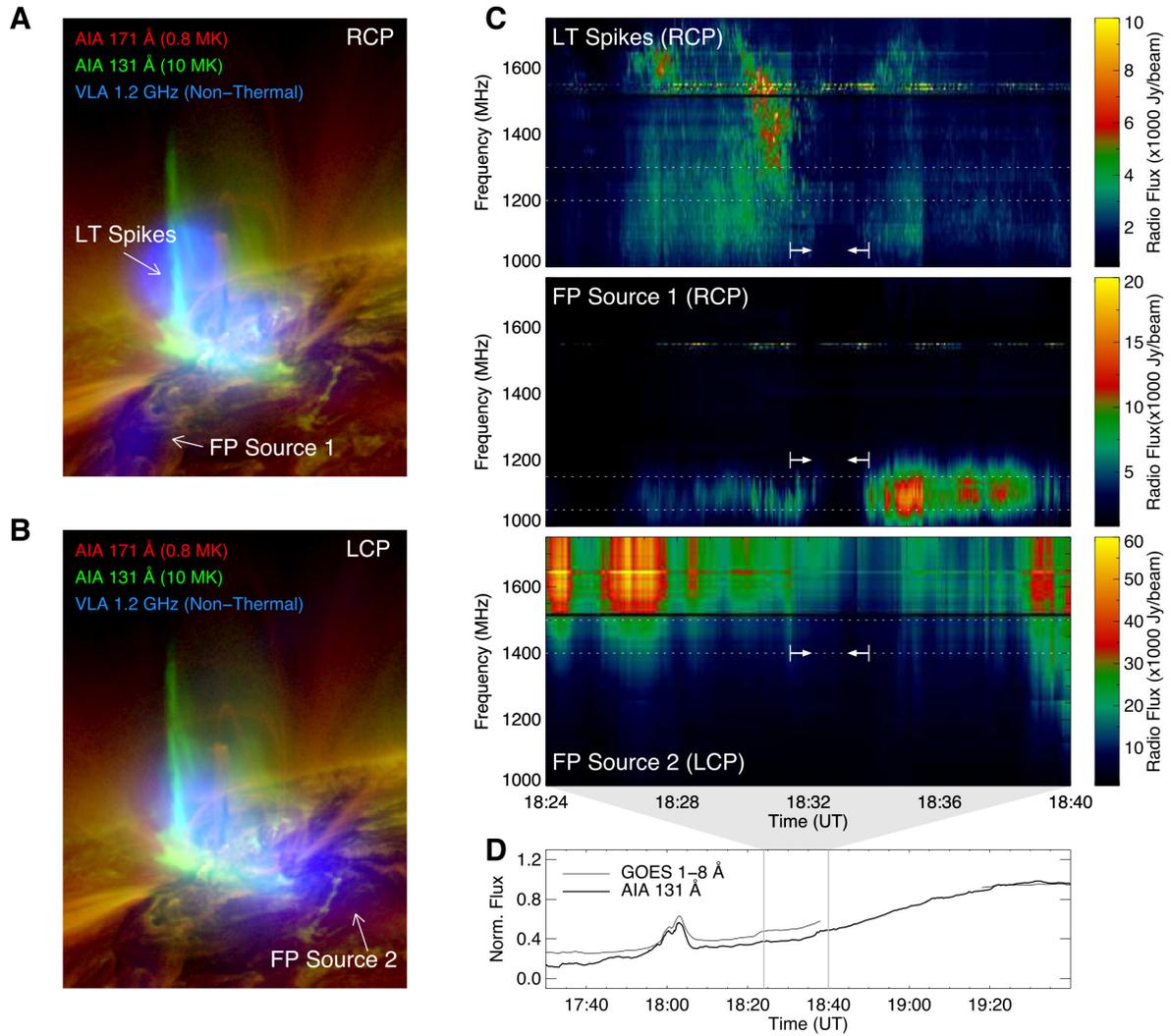

**Fig. S1. Vector dynamic spectra of spatially separated radio sources.** (**A, B**) Three radio sources present in the VLA images during the rise phase of the eruptive flare, one at the LT (labeled "LT spikes" in **A**; polarized in RCP), and two at conjugate FPs of the flaring loops (labeled "FP Source 1" and "FP Source 2" in **A** and **B**; polarized in RCP and LCP respectively). (**C**) Vector dynamic spectra for the three radio sources (each is shown in its dominant polarization). The LT source appears as the stochastic radio spikes, and the two FP sources appear as quasi-periodic pulsations. Arrows bracket a period when the TS experiences a major disruption starting from 18:31:27 UT (or $t_0 + 30$ s labeled in the last panel of Fig. 3D; same as those shown in Figs. 2D and 4). Dashed horizontal lines in each panel indicate the frequency range used to obtain the spatially resolved radio flux in Fig. 4B. (**D**) EUV and SXR light curves of the flare event.

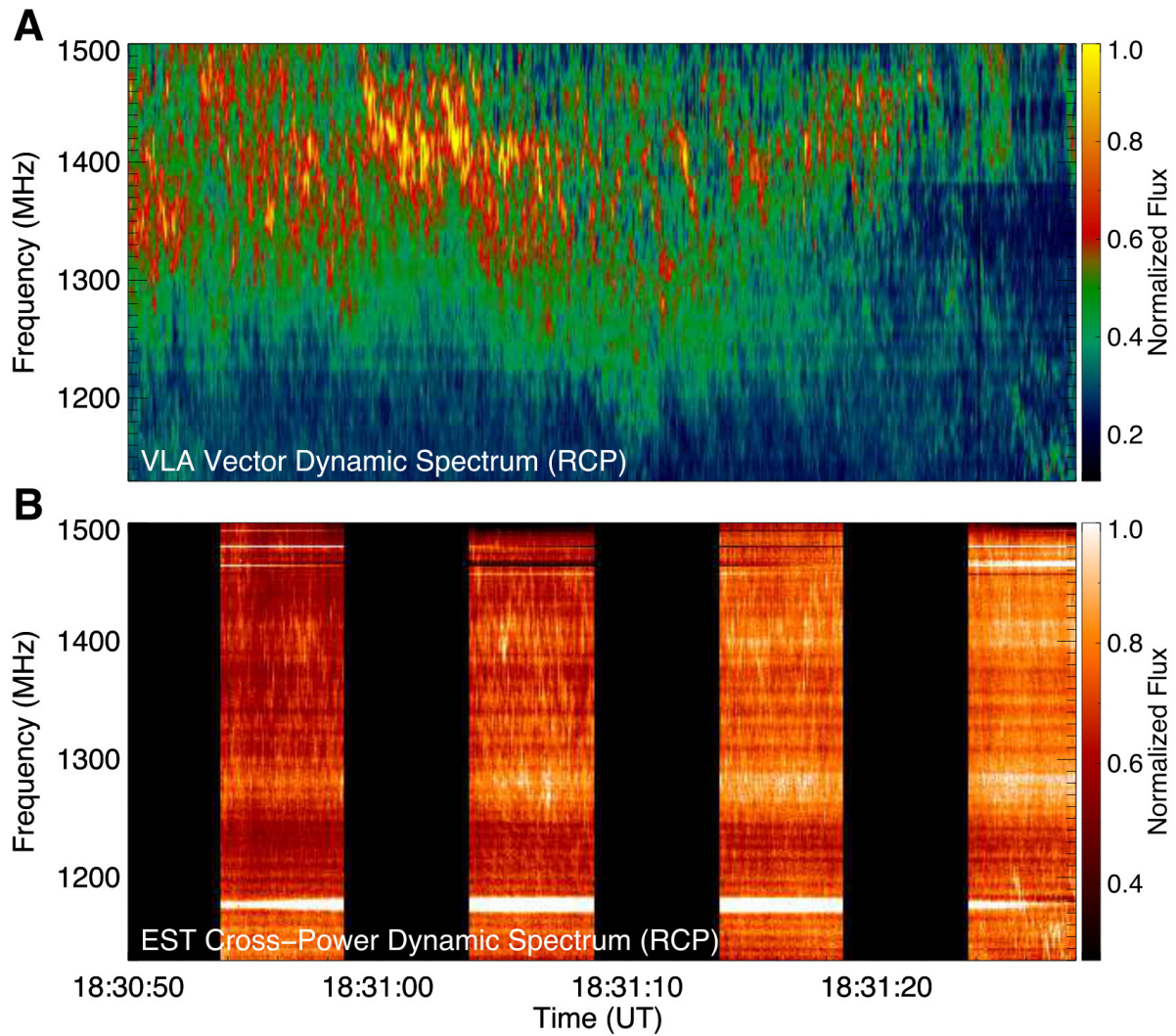

**Fig. S2. Radio dynamic spectra of the stochastic spike bursts observed by VLA and EST.**
(**A**) Vector dynamic spectrum of the spike bursts observed by the VLA in RCP (identical to Fig. 2E), which isolates the spectro-temporal information intrinsic to the spike source itself. (**B**) Cross-power dynamic spectrum (resulting from the cross-correlation of radio intensities measured at two spatially separated antennas) observed by the EST in RCP at the same time and frequency as those in (**A**). The EST dynamic spectrum shows the spike bursts at higher spectral and temporal resolution (0.5 MHz and 20 ms, respectively), but is "contaminated" by the contribution from radio sources elsewhere in the flaring region.

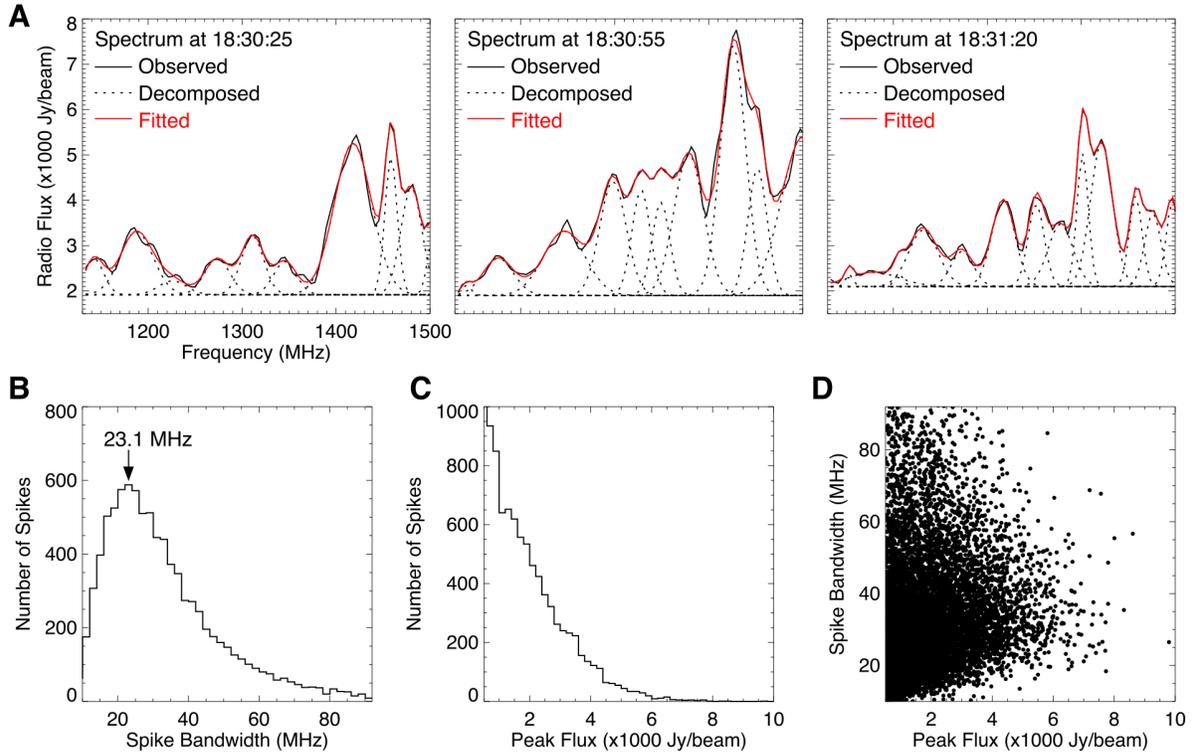

**Fig. S3. Spectral properties of stochastic spike bursts.** (**A**) Examples of the multi-Gaussian spectral fit results for VLA radio spectra at three different times. Black and red solid lines (almost indistinguishable) are the observed and fitted spectra, respectively. The decomposed Gaussian spectral profiles of the individual spike bursts are indicated as the dotted curves. (**B, C**) Bandwidth and peak flux distribution of the spike bursts. (**D**) Scatter-plot of the peak flux vs. bandwidth of the spike bursts. The statistic results are obtained from the VLA vector dynamic spectrum of the spike bursts from 18:30:20 UT to 18:31:20 UT in 1.13–1.50 GHz.

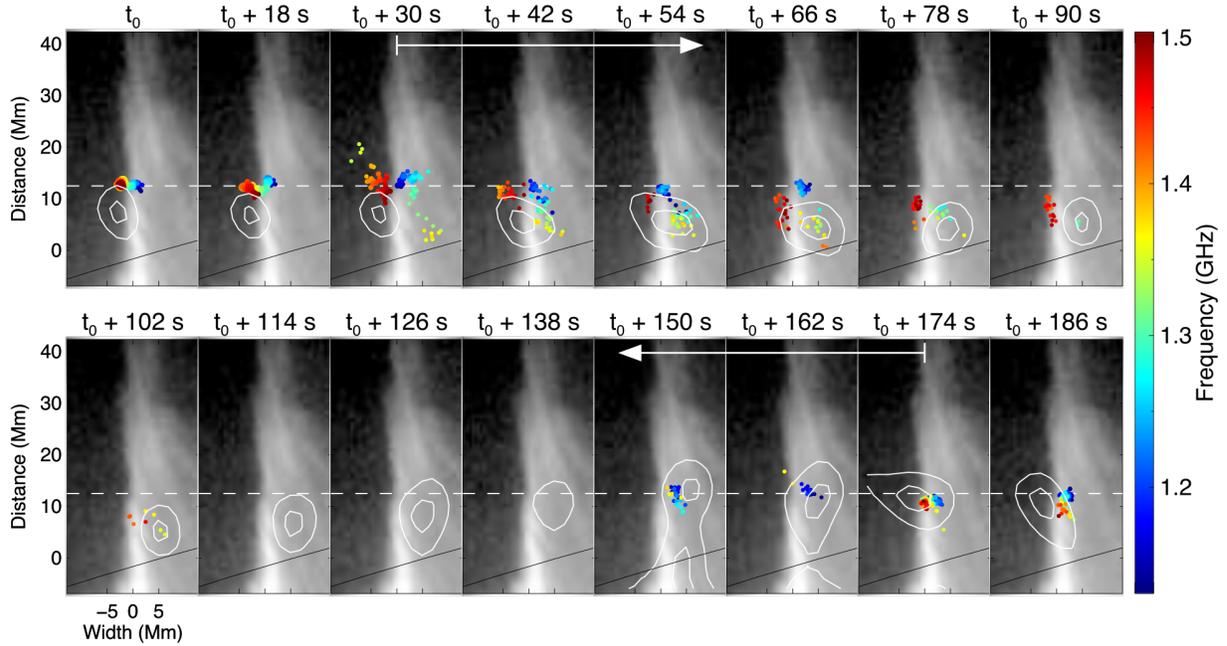

**Fig. S4. Disruption of the termination shock.** Panels increase in time from left to right and top to bottom, showing the TS as a dynamically evolving surface delineated by myriad point-like radio sources. Each colored dot corresponds to the centroid location of a radio spike at a given frequency (scaled in their flux density and colored according to the emission frequency). White contours are RHESSI 15–25 keV HXR images at the same times, made with 96-s time integration. Contour levels are 81% and 95% of their peak intensity. Background is the AIA 94 Å EUV image featuring the hot flaring magnetic loops. A pair of arrows brackets a period when the TS experiences a major disruption starting from 18:31:27 UT (or $t_0 + 30$ s labeled in the last panel of Fig. 3D; same as those shown in Figs. 2D, 4, and S1C).

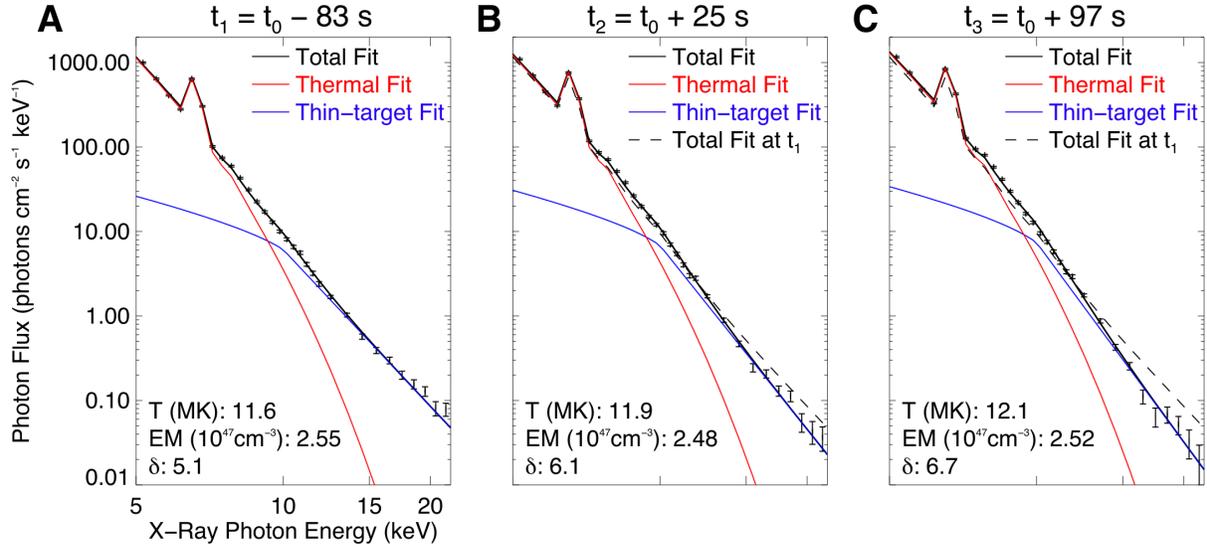

**Fig. S5. X-ray spectra before and during the termination shock disruption.** (**A**) Observed and fitted RHESSI X-ray photon spectra before the TS disruption at 18:29:34 UT (labeled as "$t_1$" in Fig. 4A). (**B, C**) RHESSI X-ray spectra during the TS disruption at 18:31:22 UT and 18:32:34 UT (labeled as "$t_2$" and "$t_3$" in Fig. 4A). Red and blue lines are the isothermal and non-thermal components of the fit results (black symbols). Black solid lines are the sum of the two components. Black dashed lines in **B** and **C** indicate the fitted spectra at the time before the TS disruption ($t_1$) for comparison. A softening electron energy spectrum during the shock disruption is evident. Note Fermi/GBM X-ray spectra also show a similar behavior.

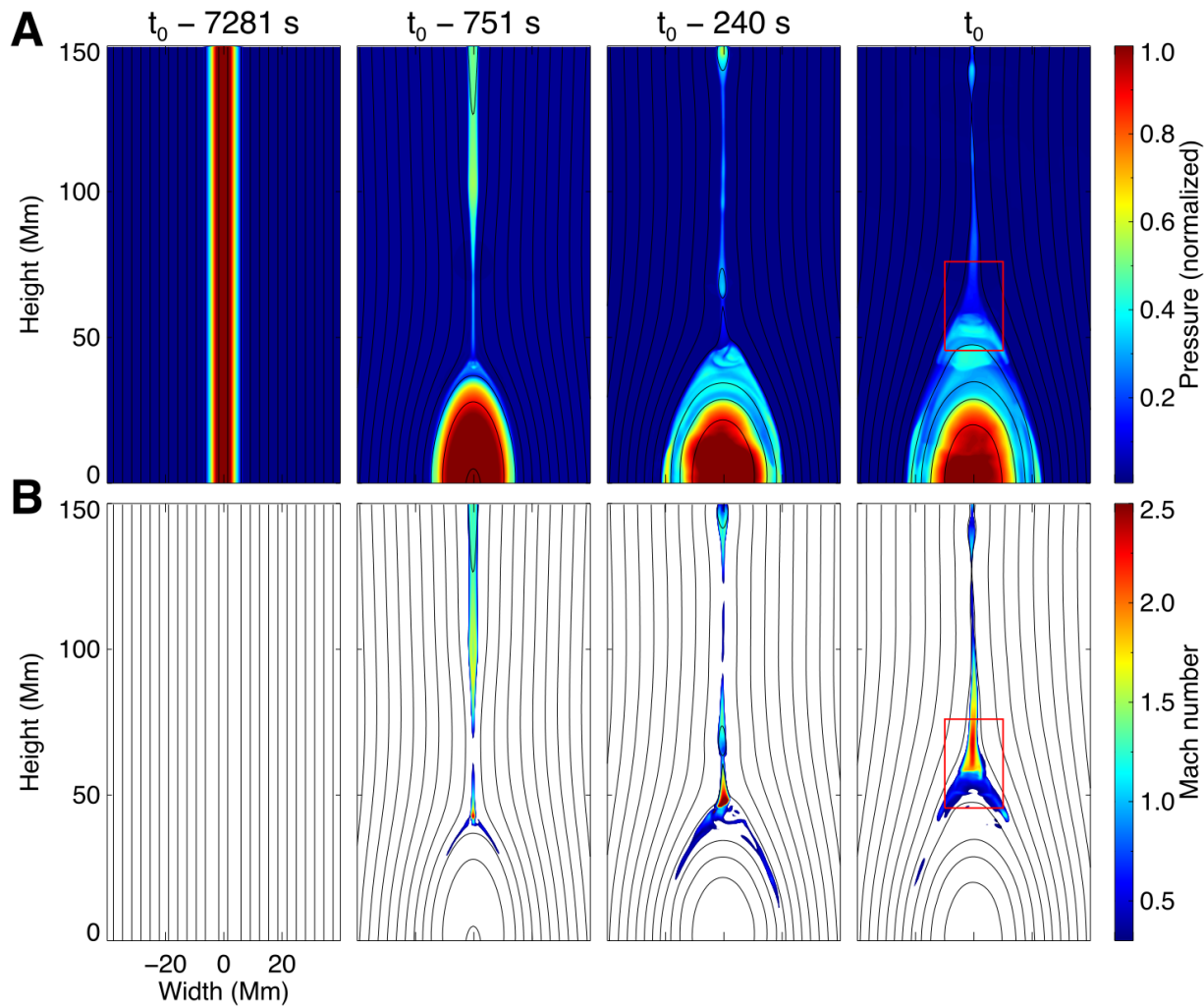

**Fig. S6. Numerical model of magnetic reconnection in the solar flare.** (**A**) and (**B**) show pressure and magnetosonic Mach number, respectively. The simulation starts from a vertical current sheet formed by anti-parallel magnetic field lines (black curves) that are tied to the photosphere (first column). It evolves self-consistently toward a cusp-shaped geometry as in the standard eruptive flare scenario and in our observations, with bi-directional reconnection outflows emanating from the reconnection site (second column). A TS is formed at LT, located below the reconnection outflows and above the newly reconnected magnetic loops (third and fourth column). The TS is visible in the simulations as a sharp layer in the last panel in (**B**) where a large velocity discontinuity presents. The labeled times are referenced to 18:30:57 UT in the observations (marked as $t_0$ in Fig. 3). The red box in the fourth column indicates the field of view of Fig. 3.

**Movie S1. Observation and simulation of the dynamic termination shock.** This animation corresponds to Fig. 3, showing observed (left panel) and simulated results (right panel) of the termination shock from 18:30:48 UT to 18:31:34 UT (or $t_0 - 9$ s to $t_0 + 37$ s). The observed shock front is delineated by colored dots. In the simulation, the shock is visible in the Mach number maps as a sharp layer of discontinuity at the front of the downward moving flow (where color changes suddenly from red/yellow to green/blue). The observed fast PD that disrupts the shock front is highlighted as a red circle (same as Fig. 3), corresponding to a rapidly contracting magnetic loop in the simulation (red curve).